\documentclass[10pt, aip, apl, twocolumn]{revtex4-1}
\usepackage{graphicx}\usepackage{amssymb}
\begin{document}
\title{Transition from columnar to point pinning in coated conductors: critical currents, that are independent of magnetic field direction }
\author{Yuri L. Zuev}
\affiliation{Department of Physics, University of Tennessee, Knoxville, TN, 37996, U.S.A.}
\author{Sung Hun Wee}
\affiliation{Oak Ridge National Laboratory, Oak Ridge, TN, 37831, U.S.A.}
\author{David K. Christen}
\affiliation{Oak Ridge National Laboratory, Oak Ridge, TN, 37831, U.S.A.}

%\author{James R. Thompson}
%\affiliation{Department of Physics, University of Tennessee, Knoxville, TN, 37996, U.S.A.}
%\affiliation{Oak Ridge National Laboratory, Oak Ridge, TN, 37831, U.S.A.}

\begin{abstract}
We identify a phase transition in the vortex system of a high-temperature superconductor with nano-columnar stacks of precipitates as strong vortex pinning centers. Above a particular, temperature-dependent field $B_X(T)$ the vortex response is no longer determined by the nano-columns, and is instead determined by point-like pinning. This phase transition leads to the change in the measured critical current density as a function of angle between the applied magnetic field and the nano-columns. Below the unbinding, there is a strong maximum in $J_C$ when field is aligned parallel to the columns. Above the unbinding, there is a minimum for this orientation.   
\end{abstract}
\maketitle

The use of high-temperature superconducting (HTSC) wires and cables in electric power industry hinges upon our ability to engineer defect structures that effectively immobilize magnetic flux lines (Abrikosov vortices) in the presence of large magnetic fields and transport current densities. Significant recent progress in this direction has been made with the discovery of nano-columnar stacks of second-phase precipitates, which form along HTSC's crystallographic $c$-axis under certain synthesis conditions~\cite{Kang, Goyal}. Such nano-columns are clearly advantageous because the vortices bind strongly to them along the entire length of the column, rather than at a few points. As a result the critical current density $J_C$ increases several-fold~\cite{Wee1}. Original discovery has been of the columns of Barium Zirconate BaZrO$_3$ (BZO), and other materials have been shown since to form similar extended structures~\cite{Varanasi, Wee2, Wee3, Horide, Harrington}. In a medium-strength magnetic field (often 1 Tesla is used) such nano-columns usually manifest themselves through a prominent maximum in angle-dependent $J_C$ when the field aligns parallel to them.
 
In addition to columnar pinning centers, the point-like pinning is also present in real superconductors. Crossover between columnar and point-like pinning has been recently addressed by, for instance,  Horide et al.~\cite{Horide1}. For HTSC materials without $c$ axis oriented columns, the $J_C$ typically has a shallow minimum when magnetic field points along $c$ axis, owing to the layered crystal structure of cuprates, captured by the Ginzburg-Landau mass anisotropy. The isotropic pinning by randomly distributed point-like defects does not modify this angular dependence appreciably. The anisotropy due to columnar pins, aligned parallel or almost parallel to crystallographic $c$-axis of the superconductor matrix, appears to compete with anisotropic superconductivity due to the layered crystal structure of the material.

In a recent paper~\cite{Zuev} we have reported the observation of an isotropic response to the applied magnetic field orientation in intrinsically anisotropic high-$T_C$ superconductor (NdBa$_2$Cu$_3$O$_7$ in that case) with columnar defects, comprised of self-assembled  BZO stacks. Specifically, the basal-plane critical current density $J_C$ becomes independent of the field orientation $\theta$ with respect to the superconductor $c$-axis over a large range of angle $\theta$. This phenomenon occurs at a particular, temperature-dependent field $B_X(T)$, where the intrinsic anisotropy of the layered high-$T_C$ superconductor and the extrinsic anisotropy, induced by the BZO stacks, effectively compensate. An analogous observation has been made by A. Johansson et al. whereby the magnetoresistance in thin films of InO$_x$ loses its orientation dependence at a particular magnetic field~\cite{Johansson}. 
\begin{figure}[b]
\centering
\includegraphics[width=\columnwidth]{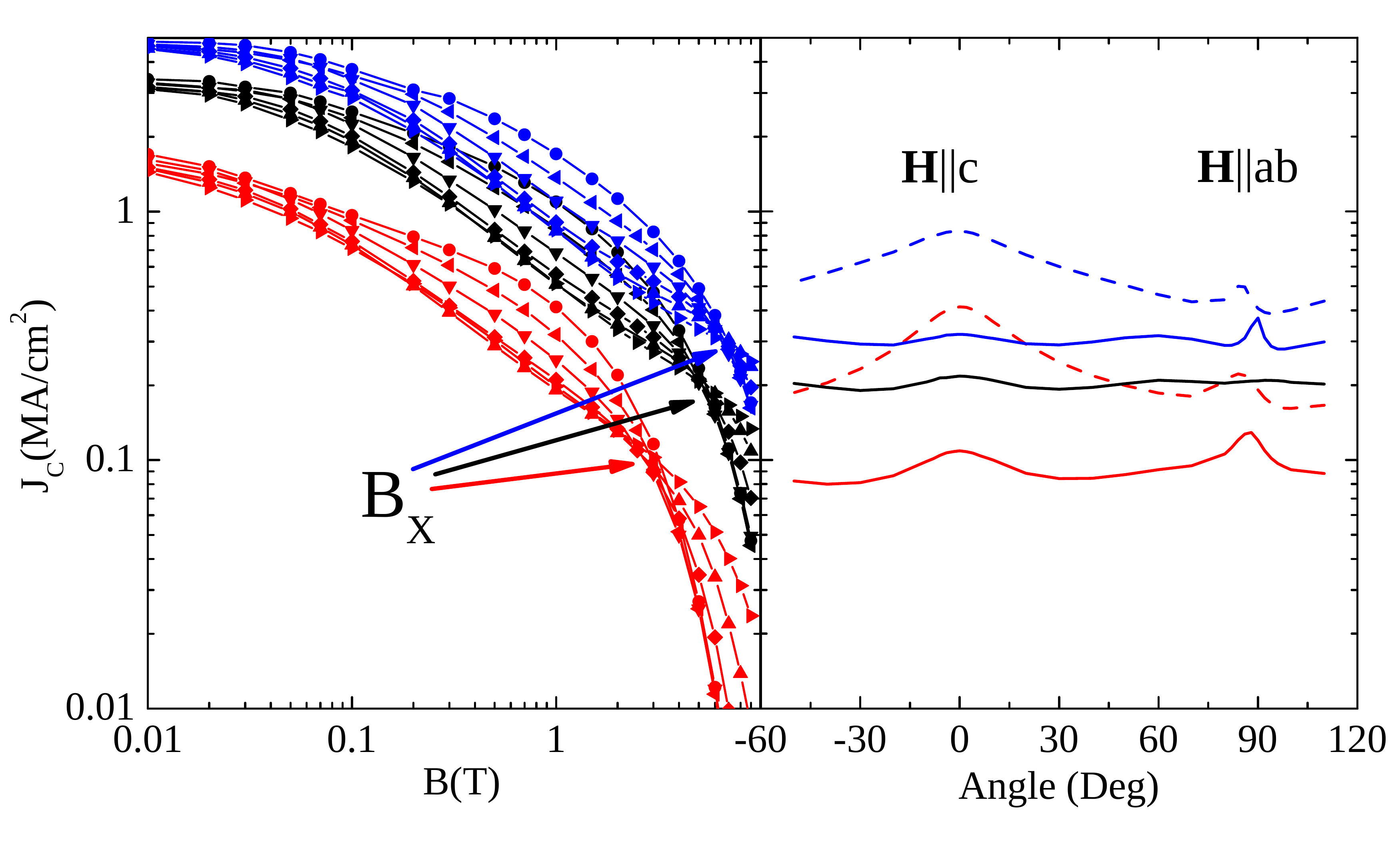}
\caption{(color online) The left panel shows three families of $J_C(B)$ curves measured at $T=$ 65 K (blue), 70 K (black) and 77 K (red), taken at different magnetic field orientation relative to the YBCO $c$-axis, from $\mathbf{H}\parallel c$ ($\circ$) to 75$^{\circ}$ ($\triangleright$). At each temperature there is a specific field $B_X$ where all the curves in the family cross. At that field and temperature the orientation dependence $J_C(\theta)$ becomes quite flat, except near $\mathbf{H}\parallel ab$ (right panel). Also shown in the right panel are the standard benchmark data at 77 K, 1 T (red dash) and 65K, 3T (blue dash), indicating presence of correlated pins aligned near $c$-axis.}
\label{fig:JcH}
\end{figure}

As a further example of such behavior, Fig.~\ref{fig:JcH} shows the $J_C$ vs. magnetic field $B$ data from a 0.8 $\mu$m thick YBCO sample deposited on an Ion-Beam Assisted Deposition (IBAD) template with columnar defects comprised of cubic double perovskite Ba$_2$YNbO$_6$ (BYNO) nanodots/rods. Details of the sample preparation and characterization can be found in ref.~\cite{Wee2}. The critical current density has been measured by a standard four-probe transport technique, maintaining $\mathbf{J}\perp \mathbf{B}$ throughout. The criterion of 1 $\mu$V/cm of dissipation has been used for $J_C$ determination.

Three families of $J_C(B)$ data, corresponding to $T=$ 65, 70 and 77 K are shown. Individual curves within each family correspond to different orientation of magnetic field with respect to sample normal. Again, we see a common crossing point $B_X$ at each temperature. The right panel of fig.~\ref{fig:JcH} shows the  $J_C$ which is roughly field orientation-independent in the angular range $-50^{\circ}\leq\theta\leq 70^{\circ}$, as measured at the respective $B_X$ at each temperature. For comparison, we also show the standard benchmark data at 77 K, 1T (red dash) and 65 K, 3T (blue dash). The strong peak near $\theta=0$ signals the presence of correlated, angle-selective pinning.

\begin{figure}[t]
\centering
\includegraphics[width=\columnwidth]{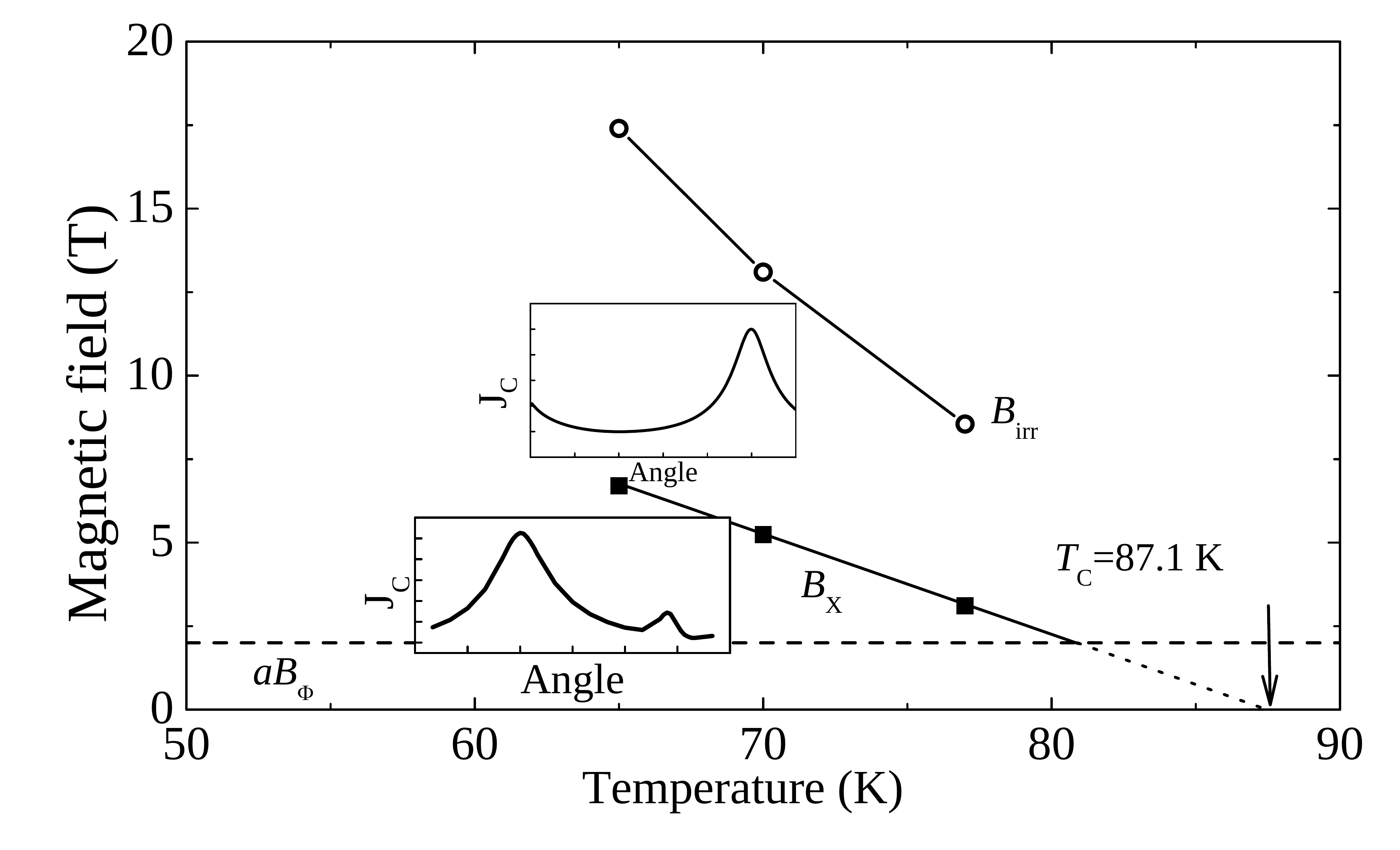}
\caption{The values of common crossing field $B_X$ ($\blacksquare$) and of the irreversibility field $\parallel c$ ($\circ$). The line $B_X(T)$ separates two types of angular dependence $J_C(\theta)$, shown schematically in the insets. The lower inset shows the actual data at 77K, 1T (see fig.~\ref{fig:JcH}), the upper inset shows a model curve. The line $B_X(T)$ extends down to a few times the matching field, $aB_{\Phi}$, shown by the dashed line, where $a$ is slightly greater than unity. The dotted line shows that $B_X(T)$ extrapolates to $T_C$.}
\label{fig:BT}
\end{figure}

The $B_X(T)$ values form a line in the $B-T$ plane, (fig.~\ref{fig:BT}, squares), which extrapolates to the $T_C$. Below and to the left of this line, the $J_C(\theta)$ response has a strong peak near $\theta=0$, corresponding to a strong $c$-axis correlated pinning, while above and to the right of this line the $J_C(\theta)$ has a minimum at $\theta=0$, which suggests isotropic weak collective pinning by point-like defects. The two insets in fig.~\ref{fig:BT} show schematic $J_C(\theta)$ behavior on either side of the phase boundary $B_X(T)$ .  We can also estimate a magnitude of the irreversibility field $B_{\mathrm{irr}}$ by fitting the $J_C(B)$ data to a phenomenological model (see~\cite{Zuev} for details). The so determined $B_{\mathrm{irr}}$ along $c$ axis are shown in fig.~\ref{fig:BT} by circles, although the conclusions of this paper are insensitive to the exact values of the irreversibility field.

The sharpness of the $B_X$ feature in fig.~\ref{fig:JcH} suggests that it is a phase transition, rather than a broad cross-over between two regimes. Below the transition the vortices are straight and bound to the columnar defects, while above the transition  the vortices unbind from the columns and meander around and between them, sampling mostly the point-like pins present in the superconducting matrix. As a possible driving force for this transition, we first consider the gain in the vortex configurational entropy
\begin{equation}
\Delta S=k_B\ln(\Phi_0/\pi B\xi_{ab}^2)^{d/\ell_z}=k_B\frac{d}{\ell_z}\ln\frac{B_{c2}}{B}
\label{eq:entropy}
\end{equation}
where the factor under the logarithm is a number of different ways to position the vortex core (of the cross-section area $\pi\xi_{ab}^2$) in an area per single vortex $\Phi_0/B$. The power $d/\ell_z$ is the number of independently pinned vortex segments of the length $\ell_z$ (the collective pinning, or Larkin, length) in a film of thickness $d$. From the critical current density we can obtain the Larkin length~\cite{Blatter} $\ell_z=\epsilon\xi_{ab}(J_{\mathrm{dp}}/J_C(0))^{1/2}\approx 1$ nm, where $\epsilon\approx 0.2$ is the YBCO mass anisotropy and the depinning current density $J_{\mathrm{dp}}$ is about ten times larger than the measured zero-field critical current density $J_C(0)$. We obtain $\Delta S\approx k_B$ per vortex per CuO$_2$ layer. This is comparable to $\approx 0.4 - 0.5\;k_B$ per vortex per layer, measured in the first-order vortex freezing transition in clean YBCO single crystals~\cite{Schilling}. However, this entropy gain must be weighted against extra energy cost of forming vortex half-loops. The appropriate energy is of the order of $\Delta U =\epsilon_1(\phi_0/B)^{1/2}$ 
where $\epsilon_1=\epsilon(\phi_0/4\pi\lambda_{ab})^2\ln{\lambda_{ab}/\xi_{ab}}$ is the vortex line energy for a vortex parallel to $ab$ planes. The energy $\Delta U$ is of the order $10^3$ K, about an order of magnitude greater than the latent heat $T_X\Delta S$ (where $T_X(B)$ is the temperature corresponding to the field $B_X$), therefore the small extra entropy can not be responsible for the discussed change in the vortex configuration. 

To seek an alternative explanation, we note that there is no sharp change in each individual $J_C(B)$ curve at the $B_X$. This may be because $B_X$ is always above the matching field $B_{\Phi}$, and therefore some vortices can not be accommodated by a strong pin. In this sample the matching field is apparently about 1 T (analysis of matching effects in materials with columnar pinning will be published elsewhere~\cite{Sinclair}). Some vortices are pinned weakly even below  $B_X$. Nevertheless, the $J_C$ does not decrease abruptly above matching. Rather, it shows a broad, $T$-independent hump, especially noticeable in the $\theta=0$ data of fig.~\ref{fig:JcH}. Smoothness across $B_{\Phi}$ may be due to significant reduction of vortex localization even below the matching field, similar to that reported in Ref.~\cite{KE}. There it was found that for magnetic fields below $B_{\Phi}$  the vortices were strongly delocalized above approximately 40 K (and independent of the areal density of the columns), producing a characteristic drop in the critical current density. 
If the vortices in our samples are not tightly bound to the columns at intermediate fields and liquid nitrogen temperatures, then as the flux density grows, the correlated nature of the pinning centers becomes screened by the ``interstitial'' vortices and the angular dependence of the critical current density changes qualitatively from that appropriate to columnar pins to the one characteristic of isotropic random pinning. A related set of ideas has been discussed in the past by Radzihovsky~\cite{Radzi}. The $1-T/T_C$ temperature dependence of the characteristic field $B_X$ at liquid nitrogen temperatures, which is apparent in fig.~\ref{fig:BT}, suggests that the intervortex distance below which the nanocolumns become screened, is given by a constant fraction of the in-plane magnetic penetration depth $\lambda_{ab}$. Given the typical values of the YBCO penetration depth of 200-250 nm at these temperatures, the screening occurs at $\sim 0.1\lambda_{ab}$, i.e. 20-25 nm. 

\begin{figure}[t]
\centering
\includegraphics[width=\columnwidth]{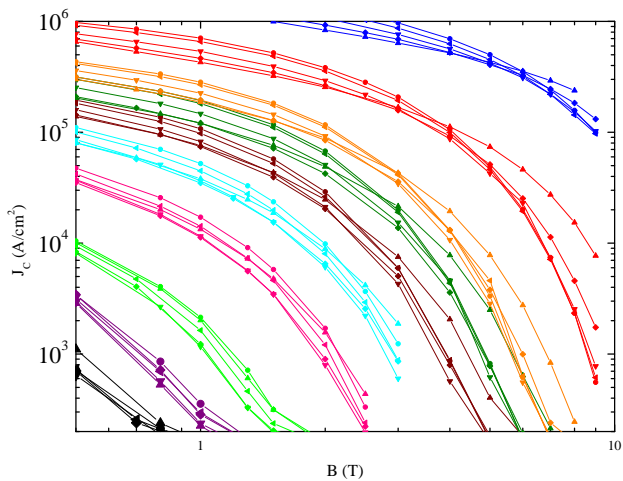}
\caption{(color online) Series of $J_C(B)$ data for a sample with Ba$_2$YZrO$_6$ nanocolumns, similar to that of fig.~\ref{fig:JcH}. The temperatures are, from top to bottom, 70, 77, 80, 81, 82, 83, 84, 85, 85.5, and 85.8 K. In each family symbols of different shape represent different field orientation:$\blacktriangle ,\;\theta=60^{\circ}$, to $\bullet, \;\theta=2^{\circ}$. Above $T=80$ K (orange) it is difficult to identify unambiguously the common crossing field $B_X$.}
\label{fig:BYZO}
\end{figure}

When the temperature rises so that $B_X$ decreases to where there are insufficient interstitial vortices to screen the columns, the sharp crossing must become smeared. This means that there must be a \textit{lower} critical field  for this transition, and it is approximately few times $B_{\Phi}$. Therefore, the $B_X(T)$ line in fig.~\ref{fig:BT} terminates at $aB_{\Phi}$, rather than extending all the way to $B=0$. % This is plausible since the global symmetry of both phases is the same. 
Figure~\ref{fig:BYZO} presents evidence in favor of this idea. Since the original sample with Ba$_2$YNbO$_6$ nanocolumns was no longer available, a YBCO sample with Ba$_2$YZrO$_6$ nanocolumns was prepared and used to examine the common crossing phenomenon at high temperature and low field. The areal density of columnar defects in this sample corresponds to the matching field of about 1 Tesla and the critical temperature is 86.2 K.

 At low temperatures (70 and 77 K, blue and red respectively) there are well defined $B_X= 5.8$ and 3.6 T respectively. As the temperature increases, the $B_X$ can not be clearly identified above approximately 80 K (orange, $B_X\approx 3$ T). Above that temperature (below that field), the individual $J_C(B)$ lines still cross, but not at a common field. This supports the argument presented above, where $a\approx 3$ in this case. The matching field may be the main parameter that controls the magnitude of $B_X$

In conclusion, we have proposed an explanation for the field-orientation independent in-plane critical current at a particular, $T$-dependent field $B_X$. The phase transition within a vortex system occurs from columnar pinned vortices at low field and/or temperature to point pinned vortices at high field and/or temperature. This phenomenon may be usefully exploited in rotating machinery (motors, generators) utilizing second generation superconductive technology.

Work at ORNL supported by US DOE.

\end{document}